\documentclass[12pt]{iopart}
\usepackage{revsymb}
\usepackage{multirow}
\usepackage{graphicx}
\usepackage{subfigure}
\usepackage{graphics}
\usepackage{psfrag}
\usepackage{iopams}
\usepackage{latexsym}

\newcommand{\ket}[1]{| #1 \rangle}
\newcommand{\bra}[1]{\langle #1 |}
\newcommand{\outerpr}[2]{|#1 \rangle \langle #2 |}

\begin{document}

\title{Entanglement measurement with discrete multiple coin quantum walks}
\author{J Endrejat, H B\"uttner}
\address{Theoretische Physik I, Universit{\"a}t Bayreuth, 
  D-95440 Bayreuth, Germany}
\ead{Jochen.Endrejat@uni-bayreuth.de}
\begin{abstract}
Within a special multi-coin quantum walk scheme we
analyze the effect of the entanglement of the initial coin state.
For states with a special entanglement structure it is shown that
this entanglement can be meausured with the mean value of the walk,
which depends on the i-concurrence of the
initial coin state. Further on the entanglement evolution is
investigated and it is shown that the symmetry of the probability
distribution is reflected by the symmetry of the entanglement distribution.
\end{abstract}
\submitto{\JPA}
\maketitle

\section{Introduction}
Quantum walks are the quantum counterparts of classical random walks.
Since they were introduced in \cite{Aharonov:93}, they were investigated to construct
the same powerful algorithms for future quantum computers as in the classical
case. An overview can be found in \cite{Kempe:03}.\\
Here we are interested in an entanglement measure and will study
a special constructed discrete quantum walk scheme
with more than one coin. We will connect the entanglement of the initial coin
state with the outcoming mean value in position space. An equation will be given
that establishes the mean value as an entanglement measure for the initial coin state.
Further we will investigate the evolution of the entanglement in coin space.
\subsection{Quantum walk scheme}
We use a simplified quantum walk scheme with $M$ coins, which is in it's base similar to
the models used by Flitney et al. \cite{Flitney:04} and Brun et al. \cite{Brun:03}.\\
The Hilbert space consists of two parts, a $(M \times 2)-$dimensional coin space
$H_C^{\otimes M}$ and a $n-$dimensional position space $H_P$.
The quantum walk scheme consists only of applying repeatedly
a special transformation operator $E$ to an inital state $ \ket{\Psi(t=0)} \equiv \ket{\Psi_0}$.
\begin{equation}
\ket{\Psi(t)} = E^t \ket{\Psi_0}  
\end{equation}
The wave function $\ket{\Psi_0}$ consists of two parts, a part 
describing the coin state and the other giving the probability
of the walker to be at a certain lattice site, $\ket{\Psi_0} = \ket{\psi_C} \otimes \ket{\psi_P}$.
The transformation operator $E$ has two parts, one for the coin tossing and one for the shifting of
the walker on the lattice. As an example we show here an operator with $M=3$ coins,
which executes the operation depending on the state of qubit 2:
\begin{equation}
\fl E= \left( \underbrace{I_C \otimes \outerpr{0}{0} \otimes I_C}_{\mbox{qubit 2}} \otimes S^{+1} +
\underbrace{I_C \otimes \outerpr{1}{1} \otimes I_C}_{\mbox{qubit 2}} \otimes S^{-1} \right)
\left( \underbrace{I_C \otimes U \otimes I_C}_{\mbox{qubit 2}} \otimes I_P \right)
\end{equation}
We use the unbiased tossing operator
\begin{equation}
U= \frac{1}{\sqrt{2}}
\left(
\begin{array}{cc}
1 & i \\
i & 1    
\end{array}
\right),
\end{equation}
which means that after one step we have the same probability for going one
step to the left or one step to the right. The shifting operators move
the quantum walker on the one-dimensional lattice, $S^{+} = \sum_{x} \ket{x+1}\bra{x}$
and $S^{-} = \sum_{x} \ket{x-1}\bra{x}$, where $x$ marks the position.
The $I_{\{P,C\}}$ are unity operators in the postition
resp. the coin subspace. One can also apply the operation to 
qubit 1 resp. qubit 3. By introducing more $I_C$
operators in coin space, one can extend the $E$ operator to more coins.
\section{Analytical evaluation}
\subsection{Fourier transformation}
We use the methods described by Nayak and Vishwanath \cite{Nayak:00}
and by Brun and coworkers \cite{Brun:03} to evaluate our scheme analytically.
After applying a discrete fourier transformation to the quantum walk scheme,
we get the $U$ operator in k-space:
\begin{equation}
U_k = \frac{1}{\sqrt{2}}
\left(
\begin{array}{cc}
e^{i k}  & i e^{i k} \\
i e^{-i k}  & e^{-i k}  
\end{array}
\right)
\end{equation}
The time evolution can then easily be
calculated by the eigenvalue decomposition of the $U_k$ operator. 
The calculation of the eigenvalues yields
$\lambda_{1,2} = \frac{1}{\sqrt{2}}(\cos k \pm i \sqrt{1+\sin^2 k})$.
If we apply a similarity transformation with 
\begin{equation}
T = 
\left(
\begin{array}{cc}
c_{+} & c_{-} \\
1 & 1  
\end{array}
\right)
\end{equation}
and $c_{\pm} = e^{i k}(\sin k \pm \sqrt{1+\sin^2 k})$
we can easily calculate the matrix product of the $U_k$ operator:
\begin{equation}
U_k^t = T 
\left(
\begin{array}{cc}
\lambda_1 & 0 \\
0 & \lambda_2
\end{array}
\right)^t
T^{-1} =
\left(
\begin{array}{cc}
a   & b \\
-b^{\ast} &   a^{\ast}
\end{array}\right)
\end{equation}
with the abbreviations $a = \cos t \theta + i \frac{\sin k}{\sqrt{1+\sin^2 k}} \sin t \theta$,
$b = \frac{i e^{i k}}{\sqrt{1+\sin^2 k}} \sin t \theta$, $\theta = \arccos \frac{\cos k}{\sqrt{2}}$,
with $|a|^2 + |b|^2 = 1$.\\
\subsection{Calculation of the moments}
Brun et al.\cite{Brun:03} derived an equation for the calculation 
of the moments of the distribution in position space
\begin{equation}\label{xmintegral}
\langle x^m  \rangle = \frac{i^m}{2 \pi}  \int_{-\pi}^{\pi} dk \; 
\langle \phi_0 | (U^{\dagger}_k)^t \;
\left[\frac{d^m}{d k^m} \; U_k^t \right] | \phi_0 \rangle
\end{equation}
which we will use to calculate the mean value ($m=1$) and the variance ($m=2$).
$\ket{\phi_0}$ is the initial state in coin space. 
We start with the simplification of the integrand
for the calculation of the first moment:
\begin{eqnarray}
\fl (U^{\dagger}_k)^t \;\frac{d}{d k} \; U_k^t =
\left(
\begin{array}{cc}
a^{\ast} & -b \\
b^{\ast} & a  
\end{array}  
\right)
\left(
\begin{array}{cc}
a' & b' \\
(-b^{\ast})' & (a^{\ast})'  
\end{array}
\right)
= \nonumber\\
\left(
\begin{array}{cc}
a' a^{\ast} + b (b^{\ast})' & a^{\ast} b' - (a^{\ast})' b\\
a' b^{\ast} - a (b^{\ast})' & a (a^{\ast})' + b' b^{\ast}  
\end{array}
\right)
:=
\left(
\begin{array}{cc}
c_1   & d_1 \\
-d_1^{\ast} & c_1^{\ast}    
\end{array}
\right)
\end{eqnarray}
We find a special property of $c_1$ since 
\begin{eqnarray}
c_1 + c_1^{\ast} &= a^{\ast} a' + b b'^{\ast}  + a'^{\ast} a + b' b^{\ast}   \nonumber\\
&= (a a^{\ast})' +  (b b^{\ast})' =(|a|^2 + |b|^2)' = 0
\end{eqnarray}
and therefore $\frac{i}{2 \pi} \int_{-\pi}^{\pi} dk \; (c_1 - c_1^{\ast}) = 0$.
In the following we will use the abbreviation
\begin{equation}
\widetilde{c_1} := \frac{i}{2 \pi} \int_{-\pi}^{\pi} dk \; c_1.
\end{equation}
From the symmetry of the integrand we can further conlude that
$\frac{i}{2 \pi} \int_{-\pi}^{\pi} dk \; \bigl( d_1 - d^{\ast}_1 \bigr) = 0$.
Our results can easily be verified with these two simplifications.
In Fig. \ref{const1} we evaluated numerically the integral
$\widetilde{c_1}$ as a function of time and made a linear regression of
the result.\\
{
\psfrag{c}[][][0.8]{$\widetilde{c_1}$}
\psfrag{t}[][][0.8]{t}
\begin{figure}
\begin{center}
\includegraphics[width=7cm]{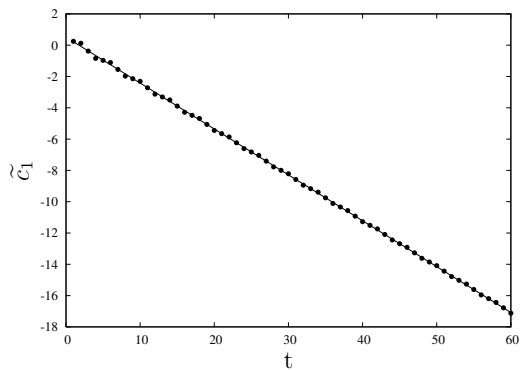}
\caption{\label{const1}The numerical results of the integral $\widetilde{c_1}$ as a function of time and
linear fit with $\widetilde{c_1}(t) = a_0 t + a_1$ with $a_0 = -0.2932$ and $a_1 = 0.5116 $.}
\end{center}
\end{figure}
}
For the calculation of the variance and of the mean deviation
one needs the second moment. For this we evaluate the integrand 
of (\ref{xmintegral}) for $m=2$.
\begin{eqnarray}
\fl (U^{\dagger}_k)^t \;\frac{d^2}{d k^2} \; U_k^t =
\left(
\begin{array}{cc}
a^{\ast} & -b \\
b^{\ast} & a  
\end{array}
\right)
\left(  
\begin{array}{cc}
a'' & b'' \\
(-b^{\ast})'' & (a^{\ast})''  
\end{array}
\right)
= \\
\left(
\begin{array}{cc}
a'' a^{\ast} + b (b^{\ast})'' & a^{\ast} b'' - (a^{\ast})'' b\\
a'' b^{\ast} - a (b^{\ast})'' & a (a^{\ast})'' + b'' b^{\ast}  
\end{array}
\right)
:=
\left(
\begin{array}{cc}
c_2   & d_2 \\
-d_2^{\ast} & c_2^{\ast}    
\end{array}
\right)
\end{eqnarray}
We know from numerical evaluation, that the integral over the matrix 
$(U^{\dagger}_k)^t \;\frac{d^2}{d k^2} \; U_k^t$ is diagonal
and that the entries have the same value:
$ -\frac{1}{2 \pi} \int_{-\pi}^{\pi} dk \; (c_2-c^{\ast}_2) = 0$
The integrals over the nondiagonal entries vanish.
With these two informations we can conclude that 
the second moment does not depend on the initial state,
but only on the integral 
\begin{equation}
\widetilde{c_2} := -\frac{1}{2 \pi} \int_{-\pi}^{\pi} dk \; c_2.
\end{equation}
This is in accordance with the results obtained by Konno \cite{Konno:02,Konno:02:1}.
In Fig. \ref{const2} we evaluated numerically the integral $\widetilde{c_2}$
as a function of time and fitted the result with a quadratic dependence.\\
{
\psfrag{c2}[][][0.8]{$\widetilde{c_2}$}
\psfrag{time}[][][0.8]{t}
\begin{figure}
\begin{center}
\includegraphics[width=7cm]{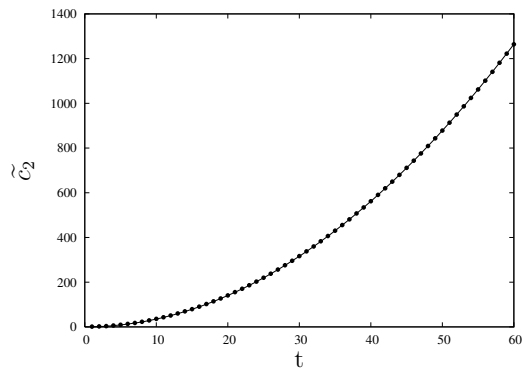}
\caption{\label{const2}The numerical results of the integral $\widetilde{c_2}$ as a function of time and
fit with $\widetilde{c_2}(t) = b_0 t^2 $ with $b_0 = 0.351249$.}
\end{center}
\end{figure}
}
We now apply our quantum walk scheme to five initial coin states,
which differ in the entanglement structure.
It is shown that for states with a certain entanglement structure,
the following equation for the squared mean value is valid:
\begin{equation}\label{mean}
\langle x \rangle^2_i = \widetilde{c_1}^2(t) \Bigl( 1-IC_i^2 \Bigr)  
\end{equation}
$\langle x \rangle_i$ is the mean value related to the
$E$ operator which acts on qubit $i$.
$IC_i$ is the i-concurrence \cite{Rungta:01} 
related to the $i$-th qubit of the initial coin state.
In addition the following equation for the variance can be derived: 
\begin{equation}\label{variance}
\left( \langle x^2 \rangle - \langle x \rangle^2 \right)_i =
\left(\widetilde{c_2}(t) - \widetilde{c_1}^2(t)\right) 
+ \widetilde{c_1}^2(t) IC_i^2  
\end{equation}
The i-concurrence measures the entanglement between two subsystems 
$A$ and $B$ and can be written as $IC_{A-B}=\sqrt{2[1-Tr(\rho^2_A)]}$,
with the reduced density matrix $\rho_A$. We use the notation $IC_{A-B} \equiv IC_A$.
Our main equation (\ref{mean}) connects the mean value of the quantum walk with this 
entanglement measure for the initial state. We show analytical and numerical examples
that eq. (\ref{mean}) only holds for so called pure entangled states, that means
states with only one kind of entanglement,e.g. for a tripartite
state either 2-qubit entanglement or 3-qubit entanglement. For states with mixed
entanglement the mean value stays zero.
\section{Application to example states}
\begin{table}
\caption{\label{structure}Entanglement structure of the example states.
2-qubit entanglement as measured by the concurrence \cite{Hill:97,Wootters:98}, 
3-qubit-entanglement for tripartite states as
measured by the tangle \cite{Coffman:00} and 3-qubit resp. 4-qubit entanglement
for fourpartite states as described in \cite{Endrejat:04}. With pure entanglement we mean, that
the state contains only one kind of entanglement.}
\begin{indented}
\item[]\begin{tabular}{ccccrc||c|c}
state & eq. & 2-qubit ent.  & 3-qubit ent.& & 4-qubit ent. & pure ent.& $\langle x \rangle_i^2 \propto IC_i^2$ \\ \hline
$\ket{\gamma GHZ}$ &(\ref{ghz})  & no  & yes &\vline& & yes & yes   \\
$\ket{\psi_6}$ &(\ref{state-psi6}) & yes & no   &\vline& & yes & yes   \\ 
$(\ket{\psi_7}+\ket{\psi_8})/\sqrt{2}$ & (\ref{psi78}) & yes & yes &\vline& & no & no  \\ \hline \hline
$\ket{\phi_1}$& (\ref{phi1}) & yes & no  && no & yes & yes \\
$\ket{\phi_2}$&(\ref{phi2}) & no  & yes?  & & yes?  & no & no
\end{tabular}
\end{indented}
\end{table}
We will apply our scheme to pure 3- and 4-qubit initial states, which differ in
their entanglement structure.
This known entanglement structure of these states is described
in \cite{Endrejat:04} and summarized in Table \ref{structure}.
\begin{table}[t]
\caption{\label{parameter}Parameter $A_0$ of the fit of the squared mean values, with
$\langle x \rangle_i^2 = A_0 (1-IC_i^2)$, for $t = 50$.
$\ket{\phi_1}_1$ resp. $\ket{\phi_1}_2$ means the state $\ket{\phi_1}$ dependent
on one parameter, when the other one is constant.}
\begin{indented}
\item[]\begin{tabular}{lrrrr}
state & $IC_1$ & $IC_2$ & $IC_3$  \vline & $IC_4$ \\ \hline
$\ket{\gamma GHZ}$   & 202.634 & 202.634 & 202.634 \vline & \\
$\ket{\psi_6}$ & 202.631 & 202.631 & 202.631 \vline &\\ \hline \hline
$\ket{\phi_1}_1$ & 202.639 & 202.636 & 202.639 & 202.634 \\
$\ket{\phi_1}_2$ & 202.641 & 202.635 & 202.641 & 202.633
\end{tabular}
\end{indented}
\end{table}
We start with a parameter dependent GHZ \cite{GHZ:89} state 
\begin{equation}\label{ghz}
\ket{\gamma\mbox{GHZ}} = \gamma \ket{000} + \sqrt{1-\gamma^2} \ket{111}  
\end{equation}
which is genuine tripartite entangled for $\gamma \in ]0,1[$.
The i-concurrence is the same for all reduced qubits,
$IC_{\{1,2,3\}}^2 = 4 \gamma^2 (1-\gamma^2)$.
The analytical solution for the mean value is found to be
\begin{equation}
\langle x \rangle^2_{\{1,2,3\}} = \widetilde{c_1}^2 (2 \gamma^2 -1)^2 
= \widetilde{c_1}^2 (1-IC_{\{1,2,3\}}^2)
\end{equation}
and we have proven the result given in (\ref{mean}).
In Table \ref{parameter} we give additionally the fit parameter $A_0$
which comes out of the fit of the simulated mean value with
$\langle x \rangle_i^2 = A_0 (1-IC_i^2)$. As one can see the simluation
and the analytical value fit very well.\\ 
The state $\ket{\psi_6}$ is a pure 2-qubit entangled parameter dependent eigenstate
of a certain spin chain and has the following form:
\begin{equation}\label{state-psi6}
\ket{\psi_6} = \kappa_1 \ket{001} + \kappa_2 \ket{010} +\kappa_1 \ket{100}    
\end{equation}
with the norm $2 \kappa_1^2 + \kappa_2^2 = 1 $ and the abbreviations
$\kappa_1 = \frac{\sqrt{\chi}}{2\sqrt{\eta}}$ and  $\kappa_2 = -\frac{2}{\sqrt{\chi}\sqrt{\eta}}$
with $\eta := \sqrt{12+\Delta(\Delta-4)}$ and $\chi := \eta + \Delta -2$. 
The i-concurrence can be calculated in terms of the parameters,
$IC_{\{1,3\}}^2 = 1-\kappa_2^4$ resp. $IC_{2}^2 = 4(\kappa_2^2-\kappa_2^4)$. 
Again it can be shown that the mean values fulfill eq.(\ref{mean}):
\begin{eqnarray}
\langle x \rangle^2_{\{1,3\}} &= \widetilde{c_1}^2  \; \kappa_2^4  
= \widetilde{c_1}^2 \Bigl(1-IC_{\{1,3\}}^2\Bigr) \\
\langle x \rangle^2_{2} &= \widetilde{c_1}^2 \Bigl(1-2\kappa_2^2\Bigr)^2 
= \widetilde{c_1}^2 \Bigl(1-IC_2^2\Bigr)     
\end{eqnarray}
{
\psfrag{delta}[][][0.8]{$\Delta$}
\psfrag{mean2}[][][0.8]{$\langle x \rangle^2_i$}
\begin{figure}
\begin{center}
\includegraphics[width=7cm]{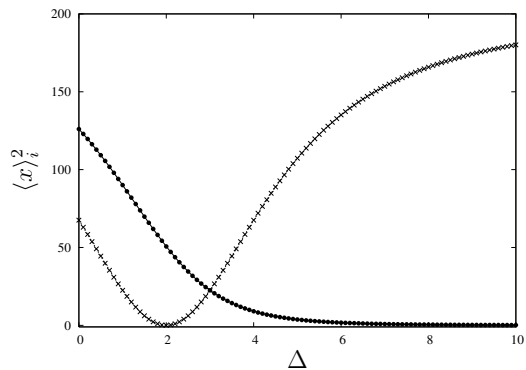}
\caption{\label{psi6}Parameter dependence of the squared mean value
for the state $\ket{\psi_6}$ for time $t = 50$. The dots mark qubit 1 resp. 3, the
x mark qubit 2. The lines are fits with $\langle x \rangle_i^2 = A_0 (1-IC_i^2)$.}
\end{center}
\end{figure}
}
In Fig. \ref{psi6} we show the parameter dependence of the squared mean value
and the result of the fit.\\
The next state at which we will have a closer look is a
superposition of two states:
\begin{eqnarray}\label{psi78}
\bigl(\ket{\psi_7} + \ket{\psi_8}\bigr)/\sqrt{2} =
\Bigl(\kappa_1 \ket{001} +  \kappa_2 \ket{010}  + \kappa_1 \ket{011} +\\
\kappa_1 \ket{100} +  \kappa_2 \ket{101}  + \kappa_1 \ket{110}  \Bigr)/\sqrt{2}
\end{eqnarray}
with abbreviations and norm from above. As we have shown in \cite{Endrejat:04},
the entanglement structure of this state shows 2-qubit as well as 3-qubit entanglement.
The i-concurrences for this state could easily be calculated,
but are not so important in this case, because the mean value
is 0, independent from any parameter:
\begin{eqnarray}
\langle x \rangle_{\{1,3\}} &=  (\widetilde{c_1}+\widetilde{c^{\ast}_1})+ 
2 \kappa_1 \kappa_2 (\widetilde{d_1} - \widetilde{d^{\ast}_1}) = 0 \nonumber \\
\langle x \rangle_2 &= ( \widetilde{c_1} + \widetilde{c_1^{\ast}})
+ 2 \kappa_1^2 (\widetilde{d_1} - \widetilde{d_1^{\ast}}) = 0 
\end{eqnarray}
These calculations are consistent with numerical simulations. Thus it
is confirmed that for mixed entanglements the mean value stays
at zero.\\
The two 4-qubit states $\ket{\phi_1}$ and $\ket{\phi_2}$
are also eigenstates of a spin chain \cite{Endrejat:04}. The parameters
$\alpha_i$ and $\beta_i$ depend on two further parameters.
The state $\ket{\phi_1}$ 
\begin{equation}\label{phi1}
\ket{\phi_1} = \alpha_1 \ket{1110} + \alpha_2 \ket{1011}+
\alpha_3 \ket{0111} - \alpha_3 \ket{1101} 
\end{equation}
with the norm $\alpha_1^2 + \alpha_2^2 + 2 \alpha_3 ^2 =1$, 
is only twopartite entangled. For simplicity we will only regard
the i-concurrence reduced on qubit 1 and 3,
$IC_{\{1,3\}}^2 = 4 (\alpha_3^2 -  \alpha_3^4 )$.
It is again nicely seen that the mean value can be described
with our equation (\ref{mean}):
\begin{equation}
\langle x \rangle^2_{\{1,3\}} = \widetilde{c_1}^2\;\Bigl(2 \alpha_3^2 -1\Bigr)^2
= \widetilde{c_1}^2\; \Bigl(1-IC_{\{1,3\}}^2\Bigr)    
\end{equation}
For the two other mean values can also be shown that they are
connected to the i-concurrences, via eq.(\ref{mean}).\\ 
The entanglement structure for the state $\ket{\phi_2}$
\begin{equation}\label{phi2}
\fl \ket{\phi_2} = -\beta_1 \ket{0011} + \beta_1 \ket{0110} - \beta_1 \ket{1001} + \\
\beta_1 \ket{1100} - \beta_2 \ket{0101} + \beta_2 \ket{1010}  
\end{equation}
with the norm $4 \beta_1^2 + 2 \beta_2^2 = 1$ is a more complex state,
since there is no possibility to show genuine 3-qubit or 4-qubit
entanglement in a fourpartite state. But in \cite{Endrejat:04} we have 
shown that this state has besides the 2-qubit entanglement either 3-qubit
or 4-qubit entanglement.
The i-concurrences reduced to one qubit are parameter 
independent, $IC_{\{1,2,3,4\}}^2 = 1$. It can be shown for this state, that
the mean values are 0,
\begin{equation}
\langle x \rangle_{\{1,2,3,4\}} = (2\beta_1^2 + \beta_2^2) 
\Bigl(\widetilde{c_1} +\widetilde{c_1^{\ast}}\Bigr) =0  
\end{equation}
These results are consistent with the 3-qubit problem and therefore
we suggest the following general result, although this is not a general proof:
For purely entangled coin-states (2-qubit, 3-qubit,...) we expect 
a square mean value proportional
to the corresponding i-concurrence, while for mixed entanglements
the squared mean value is always zero.
\section{State evolution in coin subspace}
To clear this point further we look in the following at  the entanglement evolution of the wavefunction
in the coin subspace. The complete wave function at time $t$ consists of two parts,
one part in position space and additionally a part in coin subspace,
$\ket{\Psi(t)} = \sum_x \ket{x} \otimes \ket{\psi_{\mbox{Coin}}(x)}$.
We will have a closer look at the states $\ket{\psi_{\mbox{Coin}}(x)}$.
With an artificial normalization of these states at each site one can calculate the global 
entanglement measure $Q$ \cite{MeyerWallach:02}.\\
We note the following results from our simulations: 
a) For all starting coin states the entanglement structure 
is constant in time at the starting point on the lattice;
b) at other lattice points we observe entanglement oscillations.
These two effects are shown in Fig.\ref{globaltime}, with the 
$\gamma$GHZ state as initial state. If we sit on the starting point
at $x=60$ the entanglement as measured by the global entanglement $Q$
is constant in time. If we go to $x=50$ or $x=40$ we can observe the 
described entanglement oscillations.\\
{
\psfrag{time}[][][0.8]{t}
\psfrag{Q}[][][0.8]{Q}
\begin{figure}
\begin{center}
\includegraphics[width=7cm]{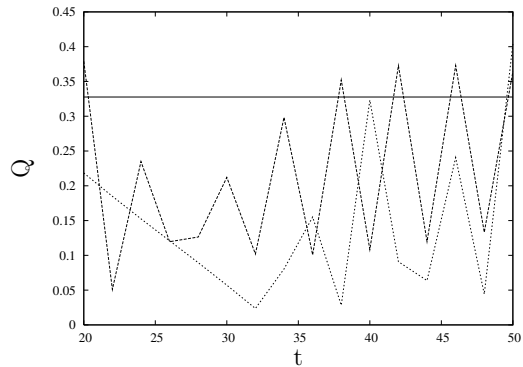}
\caption{\label{globaltime}Global entanglement $Q$ as a function of time
for the $\gamma$GHZ state as initial state, with $\gamma = 0.3$, 
viewed from different lattice points. The straight
line is for $x=60$, the dashed line for $x=50$ and the dotted line for $x=40$.
The starting point is at $x=60$.}
\end{center}
\end{figure}
}
Another effect is shown in Fig. \ref{globallattice}. The entanglement distribution
over the lattice is plotted for $t=50$ and the $\gamma$GHZ state as initial state
for two different values of $\gamma$. For $\gamma=1/\sqrt{2}$ the mean value in
position space is 0 and the entanglement distribution over the lattice is symmetric.
If we take $\gamma = 0.3$, the mean value is not equal to 0, and the entanglement
distribution is asymmetric.
{
\psfrag{site}[][][0.8]{site}
\psfrag{Q}[][][0.8]{Q}
\begin{figure}
\begin{center}
\includegraphics[width=7cm]{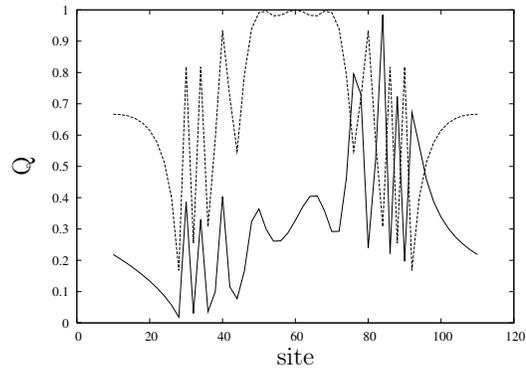}
\caption{\label{globallattice}Distribution of the global entanglement $Q$ over the lattice for $t = 50$ 
for two different $\gamma-$paramters for the $\gamma$GHZ state as initial state
with $\gamma = 1/\sqrt{2}$ (dashed line) and $\gamma = 0.3$ (straight line).}
\end{center}
\end{figure}
}
\section{Conclusions and outlook}
In conlusion, we have shown that for initial states 
with a special entanglement structure, the resulting mean value of a
multiple coin quantum walk scheme measures the entanglement of 
the initial coin state. We assume that our proposed connection
is valid for what we call pure entangled states, states with only one 
kind of entanglement.
Further on it is shown, that the symmetry of the probability distribution is
reflected by the entanglement distribution.\\
\\


\end{document}